\documentclass[journal]{IEEEtran}
\IEEEoverridecommandlockouts
\usepackage{cite}
\usepackage{url}
\usepackage{amsmath,amssymb,amsfonts}
\usepackage{graphicx}
\usepackage[caption=false,font=footnotesize]{subfig}
\usepackage{tikz}
\usetikzlibrary{positioning,arrows.meta,calc,fit,backgrounds,shapes.geometric}
\usepackage{array}
\usepackage{xcolor}
\usepackage[english]{babel}
\addto\captionsenglish{\renewcommand{\figurename}{Fig.}}
\usepackage[colorlinks=true,
            linkcolor=blue,
            citecolor=blue,
            urlcolor=blue]{hyperref}

\begin{document}

\title{DeepTelecom: A Differentiable Multimodal Framework for Channel Twinning in Dynamic Heterogeneous Wireless Environments}

\author{Bohao~Wang,
Tianen~Yuan,
William~Yen,
Zehua~Jiang,
Chongwen~Huang,
Li~Wei,
Chen~Zhu,
Jiguang~He,
Hang~Zou,
Lina~Bariah,
Brahim~Mefgouda,
Anis~Bara,
Yongliang~Shen,
Zhaoyang~Zhang,
and M\'{e}rouane~Debbah

\thanks{\textit{Corresponding author: Chongwen Huang.}}
\thanks{\textit{Part of this work was published in the proceedings of WCSP 2025~\cite{wang2025deeptelecom}.}}
\thanks{B. Wang, T. Yuan, W. Yen, Z. Jiang, C. Huang, L. Wei, and Z. Zhang are with the College of Information Science and Electronic Engineering, Zhejiang University, 310027 Hangzhou, China (e-mail: \{\href{mailto:bohaowang@zju.edu.cn}{bohaowang}, \href{mailto:yuantianen@zju.edu.cn}{yuantianen},
\href{mailto:Ycq1205@zju.edu.cn}{Ycq1205},
\href{mailto:3210104663@zju.edu.cn}{3210104663},
\href{mailto:chongwenhuang@zju.edu.cn}{chongwenhuang}, \href{mailto:l_wei@zju.edu.cn}{l\_wei}, \href{mailto:ning_ming@zju.edu.cn}{ning\_ming}\}@zju.edu.cn).}
\thanks{B. Wang, H. Zou, L. Bariah, B. Mefgouda, A. Bara, and M. Debbah are with the Research Institute for Digital Future, Khalifa University, P.O. Box 127788, Abu Dhabi, UAE (e-mail: \{\href{mailto:bohao.wang@ku.ac.ae}{bohao.wang}, \href{mailto:hang.zou@ku.ac.ae}{hang.zou}, \href{mailto:lina.bariah@ku.ac.ae}{lina.bariah},
\href{mailto:brahim.mefgouda@ku.ac.ae}{brahim.mefgouda},
\href{mailto:anis.bara@ku.ac.ae}{anis.bara},
\href{mailto:merouane.debbah@ku.ac.ae}{merouane.debbah}\}@ku.ac.ae).}
\thanks{C. Zhu is with the Polytechnic Institute, Zhejiang University, 310015 Hangzhou, China (e-mail: \href{mailto:zhuc@zju.edu.cn}{zhuc@zju.edu.cn}).}
\thanks{J. He is with the School of Computing and Information Technology, Great Bay University, Dongguan 523000, China, and Great Bay Institute for Advanced Study (GBIAS), Dongguan 523000, China (e-mail: \href{mailto:jiguang.he@gbu.edu.cn}{jiguang.he@gbu.edu.cn}).}
\thanks{Y. Shen is with the College of Computer Science and Technology, Zhejiang University, 310027 Hangzhou, China (e-mail: \href{mailto:syl@zju.edu.cn}{syl@zju.edu.cn}).}
\thanks{Data are available at \url{https://github.com/kygoyuan2004/DeepTelecom_Partial}.}
}
\maketitle

\begin{abstract}
Large-scale multimodal data are a key enabler for integrating large generative artificial intelligence models into wireless networks.
However, existing wireless datasets offer limited support for heterogeneous environments, particularly in terms of reusable high-fidelity scenes, scalable differentiable ray tracing, and synchronized multimodal data. These limitations are more pronounced in mobile and aerial scenarios with time-varying environments and channels.
To address these limitations, we propose DeepTelecom, a material-aware, high-fidelity multimodal channel dataset and reproducible generation framework for networks involving unmanned aerial vehicles and reconfigurable intelligent surfaces. DeepTelecom refines coarse indoor\,/\,outdoor reconstructions and uses a large language model to map segmented surfaces to frequency-dependent electromagnetic parameters, yielding material-aware Level of Detail 3 digital twins.
Then, a differentiable Sionna ray tracer accelerated by graphics processing units simulates propagation across the sub-$6$\,GHz\,/\,millimeter-wave\,/\,sub-terahertz bands, and specifically captures rotor-induced micro-Doppler signatures of stationary or moving unmanned aerial vehicles at different pitch angles and translational velocities.
The framework synchronously produces channel tensors, coverage maps, transmission videos, multi-view images, and weather-conditioned renderings, and supports differentiable material calibration from sparse channel anchors.
The released dataset contains $10$ material-aware indoor\,/\,outdoor scenes and $50,000$ synchronized unmanned aerial vehicle samples; the framework supports scalable generation of additional data for user-defined scenes and system configurations.
In three-dimensional unmanned aerial vehicle localization, DeepTelecom reduces the 90th-percentile error from $31.05$ meters to $2.43$ meters compared with schemes considering only channel state information, demonstrating its significance for artificial-intelligence-enabled wireless systems.
\end{abstract}


\section{Introduction}
\IEEEPARstart{T}{he} rapid progress of large multimodal models is driving the integration of artificial intelligence (AI) into sixth-generation (6G) wireless networks~\cite{ogenyi2025ainativeSurvey,zhu2025wlam,jiang2025lam,cui2025overview,li2025evolving6g,das2025emergingAI6G}. At the physical layer (PHY), large AI models are utilized to address core tasks, including multi-antenna beamforming with visual and sensor side information~\cite{roy2023multimodalbf}, channel estimation and equalization~\cite{lv2023dlchanest,zhao2024dlphy,islam2024dde2e}, massive multiple-input multiple-output (MIMO) channel state information (CSI) feedback~\cite{wen2018csinet}, radio-based sensing and localization~\cite{cha2023ailoc}, end-to-end learned transceivers~\cite{wang2018deepphy,wang2024extension}, among others. All of these tasks are data-driven. 
Their accuracy and generalization depend on datasets that jointly represent radio channels and their physical environments through multiple modalities. Hence, generating large-scale, high-fidelity, and multimodal channel data is one of the most fundamental enablers of AI-native wireless communication.

However, acquiring such datasets through real-world measurements is challenging. Channel sounding over densely sampled transmitter--receiver pairs in realistic three-dimensional (3D) environments requires costly channel sounders, robotic platforms or extensive drive tests, precise synchronization, and labor-intensive annotation. Achieving broad spatial coverage and scenario diversity is therefore prohibitively expensive~\cite{alkhateeb2023deepsense}.

To address this challenge, the research community has widely adopted digital twins based on ray tracing (RT) as the primary mechanism for generating large-scale, multimodal channel datasets~\cite{alkhateeb2019deepmimo,alrabeiah2020viwi,huangfu2022waird,shen2023dataai6g,yu2025dataaiPlus,hoydis2022sionna}. Starting from a 3D environment model, a deterministic ray tracer combines geometrical optics and electromagnetic to synthesize path-resolved channels tied to specific layouts, frequency bands, and antenna configurations. Once a scene has been modeled, RT generates controllable, repeatable channels for arbitrary virtual transmitter--receiver deployments, substantially reducing the need for exhaustive field measurements.

However, existing RT datasets remain insufficient for the demands of AI-native 6G. First, reusable scenes with consistent geometric and material fidelity are constrained. Earlier releases of DeepMIMO\,/\,ViWi mainly provided precomputed channels generated from fixed commercial RT scenarios, thereby enabling standardized learning benchmarks. However, they offered limited access to editable scene geometry, material properties, or reusable pipelines for synchronized multimodal data generation~\cite{alkhateeb2019deepmimo,alrabeiah2020viwi}. 
DeepMIMOv4 improves scene transparency and reuse by incorporating geometry, material metadata, dynamic links, and Doppler information. It also provides converters for multiple ray tracers, including Sionna RT~\cite{deepmimo2026v4}. However, it still lacks an end-to-end pipeline for constructing and calibrating material-segmented Level of Detail 3 (LoD3) indoor\,/\,outdoor twins with synchronized visual observations and wireless channel data.
More recent corpora such as WAIR-D\,/\,DataAI-6G\,/\,BUPTCMCC-6G-DataAI+ broaden system configurations and frequency coverage from sub-6\,GHz to sub-terahertz (sub-THz). However, their released scenes remain substantially coarser than material-segmented LoD3 twins and expose limited surface-level electromagnetic semantics~\cite{huangfu2022waird,shen2023dataai6g,yu2025dataaiPlus}. Reconstructing LoD3 scenes and manually assigning electromagnetic properties to every surface is labor-intensive and requires expert knowledge, limiting the scalability of material-aware digital-twin generation~\cite{gan2024quadruped,hu2023robotscan,chan2021lidarslam}. Unlike Level of Detail 1 (LoD1) block geometry, LoD3 resolves architectural elements such as walls, roofs, windows, doors, and facade structures, enabling their separation and surface-specific electromagnetic parameterization. 
RFDT-Channel integrates RGB imagery and LiDAR scans with semantic material binding and Sionna RT. However, it remains focused on site- and band-specific indoor twins rather than supporting scalable indoor and outdoor dataset families~\cite{yao2026rfdtchannel}.

The second limitation lies in integrating scalable RT with differentiable material calibration. On the one hand, graphic processing unit (GPU)-based methods built on Sionna RT, including RadioTwin and vision-language-model-guided calibration, learn material and related RF parameters from sparse measurements. However, these methods mainly target site-specific calibration or channel prediction~\cite{hoydis2022sionna,hoydis2024sionna,ait2025sionna,hoydis2024learningradio,an2025radiotwin,kang2026vlmdrt,an2026visrftwin}. On the other hand, central processing unit (CPU)-based and commercial tracers, such as NYURay\,/\,Ranplan, provide accurate propagation modeling with rich material libraries. However, their manual workflows, runtime, and access constraints limit large-scale generation~\cite{kanhere2023nyuray,ranplan2025professional}. These constraints limit differentiably calibrated channel generation across large collections of LoD3 scenes.

The third limitation is that, although recent datasets have expanded modality coverage and synchronization, their released data products remain difficult to reuse beyond the original tasks and settings. Real-world corpora such as DeepSense 6G\,/\,WiWorld-RealData\,/\,Multimodal-NF support beam prediction, localization, and low-altitude sensing. They synchronize wireless channels with camera, global positioning system (GPS), global navigation satellite system (GNSS), LiDAR, and radar data but remain tied to fixed measurement campaigns~\cite{alkhateeb2023deepsense,jiao2026wiworld,li2026multimodalnf}. Synthetic datasets such as SynthSoM\,/\,Multimodal-Wireless\,/\,SynthSoM-Twin extend these modalities with depth images, inertial measurement unit (IMU) data, weather data, and unmanned aerial vehicle (UAV) trajectories. Some also assess measurement consistency or sim-to-real transfer, but they generally release task-specific samples rather than reusable scene assets~\cite{cheng2025synthsom,mao2025multimodalwireless,chen2025synthsomtwin}. Even when UAV trajectories are provided, detailed UAV geometry, rotor motion, and the resulting micro-Doppler signatures are typically absent. Consequently, these datasets are difficult to reuse or regenerate for new carrier frequencies, array configurations, reconfigurable intelligent surface (RIS) deployments, UAV dynamics, or propagation environments.


\begin{figure*}[t]
\centering
\begin{tikzpicture}[
  font=\footnotesize,
  >={Latex[length=2mm]},
  stage/.style={rectangle, rounded corners, draw, thick, fill=blue!7, align=center, text width=2.45cm, minimum height=1.75cm, inner sep=3pt},
  outbox/.style={stage, fill=green!8},
  cal/.style={rectangle, rounded corners, draw, thick, fill=orange!15, align=center, text width=3.5cm, minimum height=1.3cm, inner sep=3pt},
  meas/.style={rectangle, draw, thick, fill=gray!12, align=center, text width=1.9cm, minimum height=1.0cm, inner sep=3pt},
  arr/.style={->, thick},
  loop/.style={->, very thick, dashed, red!70!black},
]
\node[stage] (s1) {\textbf{1) Scene Init.}\\[2pt] Outdoor 3D Tiles and indoor LiDAR\\ $\Rightarrow$ coarse geometry};
\node[stage, right=0.55cm of s1] (s2) {\textbf{2) Material Param.}\\[2pt] Blender split and LLM-assigned physical EM parameters};
\node[stage, right=0.55cm of s2] (s3) {\textbf{3) Supplement}\\[2pt] RIS, MTs,\\ and UAV motion};
\node[stage, right=0.55cm of s3] (s4) {\textbf{4) Diff. RT}\\[2pt] Sionna and OptiX\\ $\Rightarrow$ CIR/CFR, AoD/AoA, paths};
\node[outbox, right=0.55cm of s4] (s6) {\textbf{6) Visualize \& Dataset}\\[2pt] GPU videos, coverage maps;\\ multimodal HDF5/CSV};
\draw[arr] (s1)--(s2);
\draw[arr] (s2)--(s3);
\draw[arr] (s3)--(s4);
\draw[arr] (s4)--(s6);
\begin{scope}[on background layer]
  \node[draw=blue!40!gray, dashed, thick, rounded corners, fit=(s1)(s2)(s3)(s4), inner sep=9pt] (pipe) {};
\end{scope}
\node[anchor=west, font=\scriptsize\itshape, text=blue!45!black, fill=white, inner xsep=3pt, inner ysep=0.5pt] at ([xshift=12pt]pipe.north west) {Forward generation pipeline};
\node[cal, below=1.05cm of s4] (cal) {\textbf{5) Calibration loop}\\[2pt] compare simulated and target path loss\\$\nabla_{\mathbf z}\mathcal{L}_{\mathrm{total}}(\mathbf z)$ by autodiff};
\node[meas, right=0.55cm of cal] (rm) {Sparse path-loss targets};
\draw[arr] (s4) -- node[right, font=\scriptsize, pos=0.45]{sim.\ path loss} (cal);
\draw[arr] (rm) -- (cal);
\draw[loop] (cal.west) -| node[near start, below=1pt, font=\scriptsize, text=red!70!black]{update $\mathbf z$; map to $\boldsymbol{\theta}(\mathbf z)$} (s2.south);
\end{tikzpicture}
\caption{The overall DeepTelecom framework.}
\label{fig:framework}
\end{figure*}
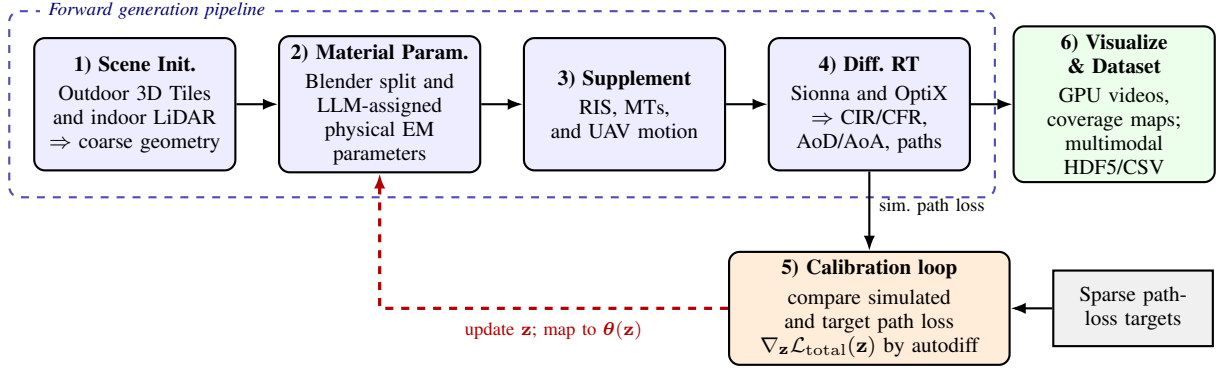

Overall, despite recent advances in standardized channel toolchains, multimodal sensing and communication datasets, RF-aware scene reconstruction, and differentiable material calibration, these capabilities remain fragmented across existing frameworks.
To bridge this gap, we introduce \emph{DeepTelecom}, a material-aware LoD3 digital-twin framework and multimodal channel dataset for 6G wireless communication. DeepTelecom constructs indoor\,/\,outdoor twins from LiDAR scans, Google 3D Tiles, procedural layouts, and existing computer-aided design (CAD) and building infor-
mation models (BIM) assets, while Blender refines their geometry and separates individual surfaces. An LLM maps the resulting semantic labels to editable, frequency-dependent permittivity, permeability, and conductivity across the sub-6,GHz\,/\,millimeter-wave (mmWave)\,/\,sub-THz bands. Built on Sionna's differentiable GPU RT engine, the propagation solver sustains tens of millions of rays per scene while supporting RISs, MTs, and UAVs~\cite{hoydis2022sionna,hoydis2024sionna}. Explicit modeling of UAV translation, pitch, and rotor motion produces rotor-induced micro-Doppler signatures beyond those of single-body UAV models. A fixed-topology differentiable calibration loop tunes material parameters from sparse path-loss measurements using robust, identifiability-aware optimization. Under a common schema, DeepTelecom synchronizes channel tensors and path-resolved tables with coverage maps, propagation videos, multi-view images, and weather-conditioned renderings.

In summary, our contributions are as follows.
\begin{itemize}
\item We develop a unified, LLM-assisted pipeline that transforms heterogeneous 3D sources into reusable material-aware LoD3 twins for indoor\,/\, outdoor environments, with editable frequency-dependent electromagnetic parameters at the surface level.
\item We build a scalable Sionna-based differentiable RT framework that sustains tens of millions of rays per scene, supports RISs, MTs, and UAVs, and models rotor-induced micro-Doppler. We exploit its differentiability through a fixed-topology, robust, identifiability-aware calibration loop that estimates material parameters from sparse channel measurements and improves path-loss and delay-spread agreement on held-out synthetic anchors.
\item We release a reusable multimodal dataset that synchronizes channel tensors, path-resolved tables, coverage maps, propagation videos, multi-view images, and weather-conditioned renderings. We demonstrate its value for UAV 3D localization, where fusing CSI with camera and path-loss images reduces the 90th-percentile error from $31.05$ to $2.43$\,m relative to CSI alone.
\end{itemize}

The rest of this paper is organized as follows. Section~\ref{sec:overview} presents the overall framework of DeepTelecom. Sections~\ref{sec:init} through~\ref{sec:packaging} detail the pipeline. 
Section~\ref{sec:experiments} simulates the dataset and analyzes the resulting channels. Section~\ref{sec:conclusion} concludes the paper and discusses future work.

\section{Framework Overview}
\label{sec:overview}

DeepTelecom turns a 3D scene into a high-fidelity, multimodal, LoD3 channel dataset through six stages, as shown in Fig.~\ref{fig:framework}. We summarize each stage below.

\begin{itemize}
    \item \textit{Stage 1, scene initialization.} We reconstruct coarse LoD3 geometry of indoor\,/\,outdoor environments from open geospatial data and LiDAR scans, and export it to industry-standard 3D formats~\cite{gan2024quadruped,hu2023robotscan,chan2021lidarslam}.
    \item \textit{Stage 2, material parameterization.} We segment the coarse scene into named surfaces and use an LLM to assign frequency-dependent electromagnetic material parameters. The output is a structured scene description in which every surface carries its material properties.
    \item \textit{Stage 3, supplementary scene elements.} On top of the parameterized geometry, we add elements that shape propagation beyond the static scene. RIS panels create controlled reflection, and MTs with velocity create time-varying channels. We further simulate the rotor-induced micro-Doppler signatures of UAVs across different pitch angles and translational velocities.
    \item \textit{Stage 4, RT and channel extraction.} We trace this fully populated scene on a GPU-accelerated, differentiable ray tracer powered by NVIDIA's Sionna and OptiX libraries~\cite{hoydis2022sionna,hoydis2024sionna,ait2025sionna}, and we derive standard channel products such as the channel impulse response (CIR), the channel frequency response (CFR), and angular statistics.
    \item \textit{Stage 5, calibration.} We calibrate the material parameters assigned in Stage 2 by minimizing a robust loss between simulated and measured path-loss values, and backpropagating its gradients through the field computation while holding the ray topology fixed.
    \item \textit{Stage 6, visualization and packaging.} After the optional calibration, we render snapshots, transmission videos, coverage maps, and weather-conditioned renderings on the GPU and organize all modalities under one unified scene schema.
\end{itemize}

\section{Initializing the Digital Twin: Coarse Scene Acquisition}
\label{sec:init}
The first stage builds the coarse LoD3 geometry that serves as the starting point of the digital twin. Two parallel pipelines produce this geometry. One reconstructs outdoor environments from open geospatial data, and the other reconstructs indoor environments from LiDAR. Both pipelines export coarse geometry through a shared interface for refinement and material parameterization in the next stage.

\subsection{Outdoor Coarse-Scene Reconstruction}
\label{subsec:outdoor}

\begin{figure}[h]
\centerline{\includegraphics[width=1\linewidth]{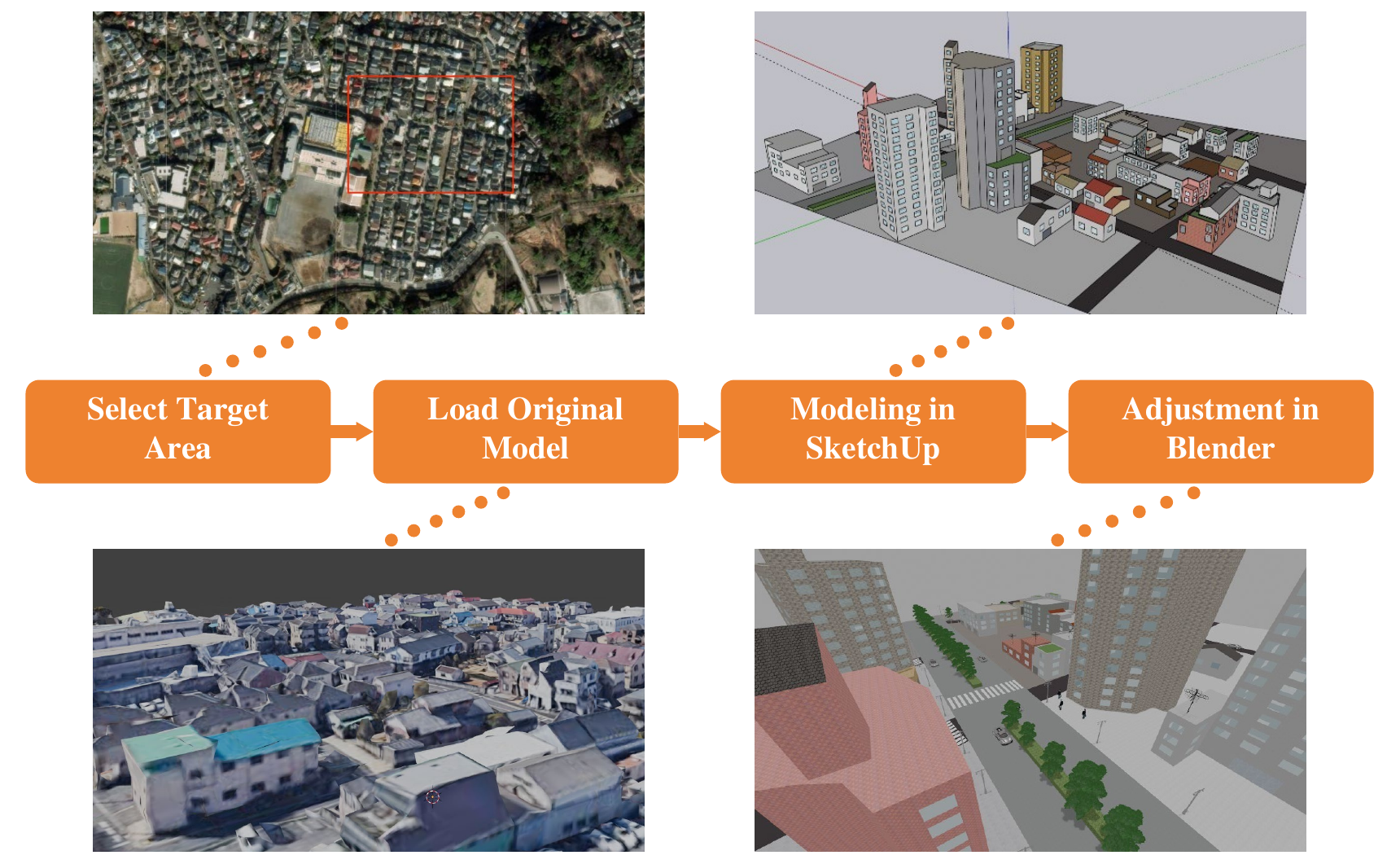}}
\caption{Outdoor digital-scene reconstruction pipeline.}
\label{fig:outdoor-scene}
\end{figure}

For outdoor environments, we build LoD3 urban scenes from open geospatial data using professional 3D modeling tools. Fig.~\ref{fig:outdoor-scene} shows the pipeline. In our implementation, the Blender-blosm plugin imports real-world models from Google 3D Tiles, our concrete source of open-access map data. A dense urban area in Japan is selected as the test site. The imported models carry building shapes, textures, and terrain elevation, but they remain coarse, with uneven surfaces and merged components. For more efficient editing, we export the models in \texttt{.gltf} format to SketchUp. In SketchUp, we rebuild the key structural elements to match the Google 3D Tiles reference, organize objects by category, and assign realistic material names such as glass and concrete. Given GPS coordinates, this pipeline reconstructs an outdoor twin with terrain and buildings.

\subsection{Indoor Coarse-Scene Reconstruction}
\label{subsec:indoor}
\begin{figure}[h]
\centerline{\includegraphics[width=1\linewidth]{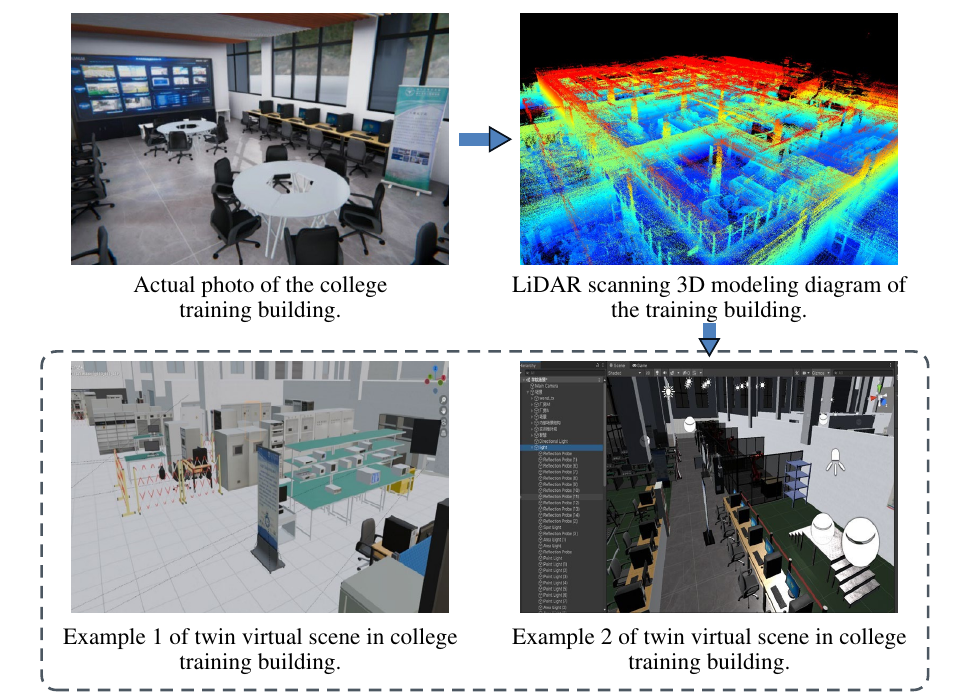}}
\caption{Indoor digital-scene reconstruction pipeline.}
\label{fig:indoor}
\end{figure}

For indoor environments, we reconstruct complex scenes at LoD3 granularity using a cyber-physical digital-twin system. A quadruped robot equipped with a 3D LiDAR scanner surveys the environment~\cite{gan2024quadruped,gan2025indoorrecon,hu2023robotscan}. The LiDAR captures both geometric distances and reflection intensities, which lets us infer surface features, such as material type and color. The raw point clouds are denoised and filtered. To transform them into a unified 3D coordinate system, we apply an HDL-localization algorithm based on an unscented Kalman filter (UKF)~\cite{zhu2024integrated}. If the UKF provides a rotation $\mathbf{R}(t)$ and a pose-translation vector $\mathbf{d}(t)$ at scan-acquisition time $t$, a point $\mathbf{p}_{\mathrm{local}}$ measured in the robot frame maps to the global frame as
\begin{equation}
\mathbf{p}_{\mathrm{global}}(t) = \mathbf{R}(t)\,\mathbf{p}_{\mathrm{local}} + \mathbf{d}(t),
\label{eq:transform}
\end{equation}
where \(\mathbf{p}_{\mathrm{global}}(t)\) is the transformed point in the global frame. This transformation aligns all scans in a common coordinate system. The aggregated point cloud can reveal the geometry and layout of indoor structures, such as walls and furniture~\cite{keller2013real,chan2021lidarslam}. Fig.~\ref{fig:indoor} shows the pipeline for indoor digital-scene reconstruction.

A 3D modeling module then converts the point cloud into a polygonal mesh. We perform mesh extraction and simplification to reduce redundant geometry while preserving essential detail, and we combine solid, surface, and wireframe modeling so that each object type receives a suitable reconstruction. We then apply texture baking and light baking, which precompute surface textures and lighting onto the mesh and yield a photorealistic appearance. The coarse indoor model is exported in standard formats such as \texttt{.obj} and \texttt{.fbx} for the next stage. Through this shared interface, indoor point clouds captured by LiDAR or red-green-blue-depth (RGB-D) sensors yield LoD3 interior models, and GPS coordinates yield outdoor twins. Both outputs feed the same parameterization stage; at this point, they contain only geometry and lack the electromagnetic semantics required for propagation.

\section{Parameterizing the Digital Twin: From Coarse Geometry to Material-Aware Surfaces}
\label{sec:param}
The second stage supplies that missing semantics by turning coarse geometry into a material-aware twin. We first segment the scene into named surfaces, and we then use an LLM to assign frequency-dependent electromagnetic parameters to each material. This material assignment differentiates DeepTelecom from nondifferentiable LoD1 corpora, which provide only extruded blocks with limited material semantics.

\subsection{Region Segmentation and Labeling via Blender}
\label{subsec:blender}
Blender unifies the indoor\,/\,outdoor pipelines on one platform. We import all reconstructed models, whether from LiDAR scans or geospatial data, using standard formats such as \texttt{.obj}, \texttt{.fbx}, and \texttt{.gltf}~\cite{hu2023robotscan}. Blender then serves three functions.
\begin{itemize}
    \item \textit{Geometric cleaning and simplification.} We repair non-manifold edges, remove duplicate vertices, and ensure watertight geometry suitable for RT. We optimize the triangle count while preserving structural detail that matters for electromagnetic propagation.
    \item \textit{Material labeling.} We assign each object a descriptive name under a strict convention, for example \texttt{wall\_concrete}, \texttt{window\_glass}, \texttt{door\_wood}, and \texttt{road\_asphalt}. This convention encodes the material into the object name and drives the next step.
    \item \textit{Texture-coordinate mapping and preservation.} We preserve texture coordinates and texture maps across import and export, which keeps scene snapshots and videos photorealistic.
\end{itemize}

After processing, each scene is exported to a custom XML representation that encodes the object hierarchy, triangle-mesh geometry, vertex coordinates, and material labels. The resulting schema is
\begin{verbatim}
<scene>
  <object id="1" name="wall_concrete">
    <mesh>
      <vertices> ... </vertices>
      <faces> ... </faces>
    </mesh>
  </object>
  ...
</scene>
\end{verbatim}
Both the LLM material module and the Sionna ray tracer parse this format.

\subsection{LLM-Assisted Material Assignment}
\label{subsec:llm}
At this stage, an LLM analyzes the structured XML scene representation to refine and validate the material annotations. This LLM-assisted material assignment process comprises four steps.
\begin{enumerate}
    \item \textit{Object name extraction.} An XML parser extracts the \texttt{name} attribute of each object, for example \texttt{window\_glass} or \texttt{wall\_brick}. These names follow the convention set during Blender modeling.
    \item \textit{LLM-based material mapping.} Using a locally deployed LLM, for example Qwen3-4B, we read each object name and query an electromagnetic material database compiled from IEEE standards and literature. The LLM performs semantic matching between the object description and the material entries. For instance, \texttt{window\_glass} maps to the dielectric parameters of soda-lime glass at the operating frequency.
    \item \textit{Parameter assignment.} For each material, we assign the electromagnetic parameters required for RT: the real relative permittivity $\varepsilon'_r$; the relative permeability $\mu_r$, which is set to one for nonmagnetic materials; the conductivity $\sigma$ in S/m; and a surface roughness coefficient for diffuse scattering. These parameters are frequency-dependent and are interpolated at the simulation center frequency. We use multilayer models for complex materials when needed.
    \item \textit{Consistency validation.} The LLM checks inconsistencies, such as adjacent objects with very different dielectric properties that would create unphysical boundaries. When it detects such cases, it suggests corrections or flags the objects for manual review.
\end{enumerate}

The material parameters enter the ray tracer through an effective
complex permittivity, Fresnel reflection coefficients, and, for rough
surfaces, a coherent roughness attenuation. Let \(s\) denote a surface
and \(m(s)\) its material class. For material class \(m\) and frequency
\(f\), under the \(e^{j2\pi f t}\) phasor convention, the effective
complex relative permittivity is~\cite{iturp2040}
\begin{equation}
\bar{\varepsilon}_{r,m}(f)
=
\varepsilon'_{r,m}(f)
-
j\frac{\sigma_m(f)}{2\pi f\varepsilon_0},
\label{eq:permittivity}
\end{equation}
where \(j=\sqrt{-1}\),
\(\varepsilon'_{r,m}(f)\) is the real part of the relative permittivity,
\(\varepsilon_0\) is the vacuum permittivity, and \(\sigma_m(f)\) is the
conductivity of material \(m\). For incidence
from free space onto a nonmagnetic locally planar surface, let
\(\theta_i\) denote the incidence angle measured from the surface normal.
The Fresnel reflection coefficients for electric-field polarizations
parallel and perpendicular to the plane of incidence, denoted by
subscripts \(\parallel\) and \(\perp\), respectively, are~\cite{iturp2040}
\begin{equation}
\begin{aligned}
\Gamma_{\parallel,m}(f,\theta_i)
&=
\frac{
\bar{\varepsilon}_{r,m}(f)\cos\theta_i
-
\sqrt{\bar{\varepsilon}_{r,m}(f)-\sin^2\theta_i}
}{
\bar{\varepsilon}_{r,m}(f)\cos\theta_i
+
\sqrt{\bar{\varepsilon}_{r,m}(f)-\sin^2\theta_i}
},\\
\Gamma_{\perp,m}(f,\theta_i)
&=
\frac{
\cos\theta_i
-
\sqrt{\bar{\varepsilon}_{r,m}(f)-\sin^2\theta_i}
}{
\cos\theta_i
+
\sqrt{\bar{\varepsilon}_{r,m}(f)-\sin^2\theta_i}
}.
\end{aligned}
\label{eq:fresnel}
\end{equation}
For a rough surface \(s\), we attenuate the coherent specular
reflection by the Rayleigh roughness factor~\cite{ament1953rough}
\begin{equation}
\rho_s(f,\theta_i)
=
\exp\!\left[
-2\left(
\frac{2\pi\sigma_{h,s}\cos\theta_i}{\lambda_0(f)}
\right)^2
\right],\;
\lambda_0(f)=\frac{c}{f},
\label{eq:roughness}
\end{equation}
where \(\sigma_{h,s}\) is the surface-height standard deviation and
\(\lambda_0(f)=c/f\) is the free-space wavelength, with \(c\) denoting
the speed of light.
The effective reflection coefficient used by the ray tracer for
polarization \(\chi\in\{\parallel,\perp\}\) is then~\cite{ament1953rough}
\begin{equation}
\widetilde{\Gamma}_{\chi,s}(f,\theta_i)
=
\rho_s(f,\theta_i)\,
\Gamma_{\chi,m(s)}(f,\theta_i).
\label{eq:effective_reflection}
\end{equation}

\section{Supplementary Scene Elements: RIS, Mobility, and UAV}
\label{sec:supplement}
Beyond static geometry and materials, DeepTelecom adds three elements that shape propagation in realistic deployments. RIS panels introduce controlled reflection, MTs introduce translational motion that makes the channel time-varying, and rotating UAVs introduce a micro-Doppler modulation on top of that motion. We configure all three before RT, and they reshape the channels the engine produces.

\subsection{RIS}
\label{subsec:ris}
We first add RIS panels, which redirect signal reflection through programmable phase control. We place each panel in the scene and support two beamforming modes.
\begin{itemize}
    \item \textit{Single-beam focusing.} For a point-to-point high-throughput backhaul link, we apply a phase-gradient strategy based on path-length differences. To steer a beam toward a direction $(\theta_0,\phi_0)$ with the RIS in the $xy$-plane, the phase profile is
    \begin{equation}
    \Phi(x,y) = -\frac{2\pi}{\lambda}\bigl(x\sin\theta_0\cos\phi_0 + y\sin\theta_0\sin\phi_0\bigr).
    \label{eq:ris_single}
    \end{equation}
    Here, $(x,y)$ are coordinates on the RIS, $\lambda$ is the wavelength,
    and $\theta_0$ and $\phi_0$ are the target elevation and azimuth angles,
    respectively. This linear phase gradient across the surface steers the
    beam toward $(\theta_0,\phi_0)$.
    \item \textit{Multibeam optimization.} For scenarios that serve multiple users simultaneously, we design phase codebooks using iterative methods such as stochastic gradient descent~\cite{zhou2021stochastic} so that the radiation pattern approximates the desired multibeam shape.
\end{itemize}

\subsection{MT Mobility}
\label{subsec:mobility}
We assign each MT a velocity vector, which makes the channel response vary as the terminal moves and allows us to generate a time-varying CIR. We add time-varying noise so that the dataset captures realistic channel dynamics. Each moving path also acquires a Doppler shift,
\begin{equation}
f_{\mathrm{D},\ell} = \frac{v}{\lambda}\cos\psi_\ell,
\label{eq:doppler_shift}
\end{equation}
where $\ell\in\{1,\ldots,L\}$ indexes the $L$ propagation paths retained for the considered link, $v$ is the relative speed, $\lambda$ is the wavelength, and $\psi_\ell$ is the angle between the path direction and the velocity vector. The ray tracer in Section~\ref{sec:raytracing} records this shift per path, and the second-order statistics in Section~\ref{subsec:products} summarize it as a Doppler spread. This per-path shift captures translation, but it cannot represent the faster modulation that a spinning rotor imposes, which the next element addresses.

\subsection{UAV Modeling and Micro-Doppler}
\label{subsec:uav}
The UAV flight state couples translational velocity, pitch angle, and rotor frequency, which affect different parts of its Doppler spectrum. Equation~\eqref{eq:doppler_shift} describes the one-way communication-link Doppler shift. The UAV sensing return considered below is monostatic, so the round-trip propagation phase doubles the translational shift. For an angle $\psi$ between the velocity vector and the line of sight (LoS), we denote the echo-center shift by $f_{\mathrm{D}}^{\mathrm{mono}}=(2v/\lambda)\cos\psi$. The pitch angle rotates the airframe and rotor plane relative to the LoS, whereas the rotor frequency $f_{\mathrm{rot}}$ sets the periodic blade motion. Together, these variables determine the center, extent, and periodic structure of the rotor-induced micro-Doppler signature (MDS).

We represent the UAV body by one point scatterer and discretize each rotor blade into points that rotate about the hub. A blade point at radius $\rho$ turns at angular rate $\omega=2\pi f_{\mathrm{rot}}$. Its projected velocity is sinusoidal, which produces the instantaneous micro-Doppler shift~\cite{chen2006microdoppler,li2025microdoppler}
\begin{equation}
f_{\mathrm{mD}}(t) = \frac{2\omega\rho}{\lambda}\cos\theta\,\cos\!\bigl(\omega t + \phi_0 - \phi_{\mathrm{LoS}}\bigr),
\label{eq:md_inst}
\end{equation}
where $t$ is continuous observation time, $\lambda$ is the wavelength, $\theta$ is the LoS elevation/look angle measured from the rotor plane, $\phi_0$ is the initial blade phase, and $\phi_{\mathrm{LoS}}$ is the LoS azimuth in the rotation plane. Equivalently, the angle between the LoS and the rotor axis is $\pi/2-\theta$. The pitch angle is distinct from $\theta$. Pitching the airframe rotates the rotor plane and axis in the world frame and thereby changes $\theta$ for the current geometry of the UAV and base station (BS). At the blade tip $\rho=R$, the maximum shift is
\begin{equation}
f_{\mathrm{mD}}^{\max} = \frac{4\pi f_{\mathrm{rot}} R}{\lambda}\cos\theta,
\label{eq:md_peak}
\end{equation}
where $R$ is the blade radius and $f_c$ is the carrier frequency in
$\lambda=c/f_c$. The spectral extent therefore grows linearly with
$f_{\mathrm{rot}}$, $R$, and $f_c$, and it is scaled by the
attitude-dependent projection $\cos\theta$.

Let $x(t)$ denote the complex return of one blade. Applying the Jacobi--Anger identity~\cite{chen2006microdoppler,li2025microdoppler} to its periodic phase modulation yields the discrete Doppler sideband expansion
\begin{equation}
x(t) = A\!\sum_{\nu=-\infty}^{\infty}\!J_\nu(\beta)\,e^{j[2\pi(f_{\mathrm{D}}^{\mathrm{mono}}\!+\!\nu f_{\mathrm{rot}})t\!+\!\nu(\phi_0\!-\!\phi_{\mathrm{LoS}})]},
\label{eq:md_bessel}
\end{equation}
where $A$ is the echo amplitude, $\nu\in\mathbb{Z}$ is the sideband order,
$J_\nu$ is the $\nu$-th order Bessel function, and
$\beta=4\pi R\cos\theta/\lambda$ is the modulation index. The body speed shifts the monostatic comb center through $f_{\mathrm{D}}^{\mathrm{mono}}$, the rotor frequency sets the single-blade spacing $f_{\mathrm{rot}}$, and the pitch-dependent $\theta$ controls the envelope through $J_\nu(\beta)$. For $B$ equally spaced blades, coherent addition retains orders separated by $Bf_{\mathrm{rot}}$, tying the observed periodicity to the blade count.

DeepTelecom generates the MDS inside the differentiable ray tracer of Section~\ref{sec:raytracing}. Let $\mathcal K$ denote the finite set of UAV point scatterers, and let $k\in\mathcal K$ index one scatterer. At each rotor snapshot, the ray tracer updates the body and blade points, obtains the one-way complex channel coefficient $h_k(t)$ from the BS to scatterer $k$, and forms the approximate monostatic echo $s(t)$ following RT-based UAV micro-Doppler modeling~\cite{li2025microdoppler}
\begin{equation}
s(t) = \sum_{k\in\mathcal K} w_k\,h_k(t)^2 + \eta(t),
\label{eq:md_echo}
\end{equation}
where $w_k$ is a dimensionless scatterer weight and $\eta(t)$ is additive complex noise. Squaring $h_k(t)$ approximates the round trip from the BS to the scatterer and back. A short-time Fourier transform of the resulting snapshot sequence yields the MDS while retaining the multipath effects of the surrounding scene.

\section{RT Engine and Channel Extraction}
\label{sec:raytracing}
With the scene parameterized and the supplementary elements of Section~\ref{sec:supplement} in place, the fourth stage traces the propagation and extracts channel products. We build on NVIDIA's Sionna RT, which combines high-performance RT with full differentiability and therefore enables gradient-based optimization of scene parameters~\cite{hoydis2022sionna,hoydis2024sionna,ait2025sionna}. The tracer operates on the detailed 3D scene together with its transmitters, receivers, RIS panels, and MTs, so a single pass propagates the signals and captures the multipath effects that these elements induce.

\subsection{RT}
\label{subsec:tracing}
Tracing begins from the devices that the earlier stages placed. Each BS and MT carries a position, an orientation, a transmit power, and an antenna array, whereas the system bandwidth sets the channel delay resolution. From every transmitter, we launch many rays with Fibonacci sphere sampling for near-uniform coverage, and we add adaptive importance sampling so that regions likely to contain receivers receive denser rays. NVIDIA OptiX traces large batches of rays in parallel and tests each one against the geometry, which sustains tens of millions of rays per scene without impractical delays~\cite{hoydis2024sionna}. We terminate a ray once it exceeds a maximum interaction depth or falls below a power threshold, and we discard duplicate paths that repeat the same interaction sequence. At each interaction, we apply the physical model and update the complex amplitude. Reflections and transmissions follow geometrical optics with the Fresnel coefficients of~\eqref{eq:fresnel} and the material permittivity of~\eqref{eq:permittivity}; diffraction follows the uniform theory of diffraction; and free-space propagation accounts for the remaining loss. For a retained path topology, Sionna RT differentiates the material-dependent field computation with respect to continuous scene parameters. Section~\ref{sec:calibration} uses this derivative for calibration.

\subsection{Channel Data Products}
\label{subsec:products}
The differentiable tracer yields not only the channel but a structured record of how each signal reaches its receiver. For each transmitter and receiver, the engine records the multihop propagation paths as sequences of coordinate points with interaction labels. Each path carries its vertices, interaction types, material identifiers, complex gain $\alpha_\ell$, delay $\tau_\ell$, angle of arrival $\Omega_{r,\ell}$, angle of departure $\Omega_{t,\ell}$, and the Doppler shift of~\eqref{eq:doppler_shift}. From these paths we synthesize the standard channel responses. At physical observation time $t$, let $\mathcal P(t)=\{1,\ldots,L(t)\}$ denote the retained path set and let $\tau$ denote the excess-delay variable. For a MIMO system, the time-varying CIR is
\begin{equation}
\begin{aligned}
\mathbf{H}_{\mathrm{CIR}}(t,\tau)
={}&\sum_{\ell\in\mathcal P(t)} \alpha_\ell(t)\,
\mathbf{a}_r\!\left(\Omega_{r,\ell}(t)\right) \\
&\quad\times
\mathbf{a}_t^{\mathsf{H}}\!\left(\Omega_{t,\ell}(t)\right)
\delta_{\mathrm D}\!\left(\tau-\tau_\ell(t)\right),
\end{aligned}
\label{eq:cir}
\end{equation}
where $\mathbf{a}_r$ and $\mathbf{a}_t$ are the receive and transmit array responses, $(\cdot)^{\mathsf H}$ denotes the Hermitian transpose, and $\delta_{\mathrm D}(\cdot)$ is the Dirac delta. The corresponding CFR is the Fourier transform of~\eqref{eq:cir} with respect to $\tau$,
\begin{equation}
\mathbf{H}_{\mathrm{CFR}}(t,f)
=
\int_{-\infty}^{\infty}
\mathbf{H}_{\mathrm{CIR}}(t,\tau)
e^{-j2\pi f\tau}\,\mathrm{d}\tau,
\label{eq:cfr}
\end{equation}
where $f$ is frequency. Thus, $t$ identifies when the channel is observed, whereas $\tau$ is solely the CIR delay coordinate. The scalar $h_k(t)$ used in~\eqref{eq:md_echo} is the selected BS-to-scatterer coefficient of $\mathbf H_{\mathrm{CFR}}(t,f_c)$ at the carrier frequency. We also compute second-order statistics of channel dispersion. At a fixed $t$, we suppress explicit time dependence and write $L=L(t)$. The root-mean-square (RMS) delay spread $\tau_{\mathrm{rms}}$ is
\begin{equation}
\tau_{\mathrm{rms}} = \sqrt{\frac{\sum_{\ell=1}^{L} |\alpha_\ell|^2 \tau_\ell^2}{\sum_{\ell=1}^{L} |\alpha_\ell|^2} - \left(\frac{\sum_{\ell=1}^{L} |\alpha_\ell|^2 \tau_\ell}{\sum_{\ell=1}^{L} |\alpha_\ell|^2}\right)^2},
\label{eq:delay_spread}
\end{equation}
and the angular spread $\sigma_{\mathrm{AS}}$ for departure and arrival is
\begin{equation}
\sigma_{\mathrm{AS}} = \sqrt{\frac{\sum_{\ell=1}^{L} |\alpha_\ell|^2 \Delta\phi_\ell^2}{\sum_{\ell=1}^{L} |\alpha_\ell|^2}},
\label{eq:angular_spread}
\end{equation}
where $\Delta\phi_\ell$ is the angular deviation from the mean direction, and the Doppler spread takes the same power-weighted form over the per-path Doppler shifts. On a specified plane, we also rasterize a numeric coverage grid whose received power $P_{\mathrm{rx}}(\mathbf r)$ follows the link budget
\begin{equation}
P_{\mathrm{rx}}(\mathbf{r}) = P_{\mathrm{tx}} + G_{\mathrm{tx}} + G_{\mathrm{rx}} - \mathrm{PL}(\mathbf{r}) \quad \text{[dBm]},
\label{eq:power_map}
\end{equation}
where $\mathbf r$ is a coverage-grid location, $P_{\mathrm{tx}}$ is in dBm, $G_{\mathrm{tx}}$ and $G_{\mathrm{rx}}$ are in dBi, and $\mathrm{PL}(\mathbf{r})$ is the path loss in dB aggregated from all paths that reach $\mathbf{r}$. The visualization stage renders this grid. After filtering unreachable paths and invalid links, DeepTelecom links the scene geometry and materials to the channel and gives a physically grounded basis for analyzing MIMO behavior in time and frequency. These products inherit the material-table initialization, which the next stage can calibrate against sparse path-loss targets.

\section{Calibration}
\label{sec:calibration}

Material tables provide physically meaningful initial parameter values, but these values may not accurately reproduce the target channel response. Therefore, we formulate calibration as a fixed-topology differentiable inverse problem. Given sparse target path-loss anchors, the method replays the material-dependent fields on retained propagation paths, minimizes a robust mismatch, and diagnoses whether the selected anchors constrain the trainable parameters. The targets can be co-registered measurements when available.

\subsection{Fixed-Topology Differentiable Forward Model}
\label{subsec:cal_problem}

At the calibration carrier frequency $f_c$, expressed in hertz, we evaluate the general material model in~\eqref{eq:permittivity} at $f=f_c$. For brevity, we suppress the frequency arguments and write $\varepsilon_{r,m}=\varepsilon'_{r,m}(f_c)$, while $\sigma_m$ denotes the conductivity evaluated at $f_c$. Updating the real quantities $\varepsilon_{r,m}$ and $\sigma_m$ changes the material-dependent Fresnel coefficients evaluated by the ray tracer.

Let $\mathcal I$ denote the set of calibration anchors, with $i\in\mathcal I$. In the differentiable RT formulation described in~\cite{hoydis2024sionna,hoydis2024learningradio}, path discovery and field evaluation play different roles. For anchor $i$ at receiver position $\mathbf{r}_i$, the path solver first finds a discrete path-index set $\mathcal P_i$,
\begin{equation}
\mathcal{P}_i
=
\operatorname{TracePaths}
\bigl(\mathcal{G},\mathbf{p}_{\mathrm{tx}},\mathbf{r}_i\bigr),
\label{eq:tracepaths}
\end{equation}
where $\operatorname{TracePaths}(\cdot)$ is the discrete path-discovery operator, $\mathcal{G}$ denotes the scene geometry, and $\mathbf{p}_{\mathrm{tx}}$ is the transmitter position. Visibility tests and path selection are discrete. Let $\mathbf z$ denote the unconstrained calibration variables and let $\boldsymbol\theta(\mathbf z)$ denote the corresponding material parameters, with the detailed mapping specified in~\eqref{eq:reparam}. We therefore trace $\mathcal{P}_i$ first and differentiate only the continuous field replay on that fixed topology,
\begin{equation}
\bigl\{\alpha_{i\ell}(\boldsymbol{\theta}(\mathbf z)),\tau_{i\ell}\bigr\}_{\ell\in\mathcal{P}_i}
=
\mathcal{F}\bigl(\mathcal{P}_i,\boldsymbol{\theta}(\mathbf z)\bigr),
\label{eq:fieldcomp}
\end{equation}
where $\mathcal{F}$ is the differentiable field-replay operator, $\ell\in\mathcal P_i$ indexes a retained path, $\alpha_{i\ell}$ is its normalized dimensionless complex channel coefficient, and $\tau_{i\ell}$ is its delay. With geometry and topology fixed, the delays remain unchanged, whereas the material interactions reweight the path coefficients. We use the phase-robust path-gain observable $\widehat g_i(\boldsymbol\theta)$,
\begin{equation}
\widehat{g}_i(\boldsymbol{\theta})
=
\sum_{\ell\in\mathcal{P}_i}
\left|\alpha_{i\ell}(\boldsymbol{\theta})\right|^2+\epsilon_g,
\label{eq:cal_power}
\end{equation}
where the normalized gain $\widehat g_i$ and stabilizer $\epsilon_g>0$ are dimensionless. The predicted path loss $\widehat{\mathrm{PL}}_i(\boldsymbol\theta)$ is
\begin{equation}
\widehat{\mathrm{PL}}_i(\boldsymbol{\theta})
=
C_i-10\log_{10}\widehat{g}_i(\boldsymbol{\theta}),
\label{eq:cal_pl}
\end{equation}
where $C_i$ is the known fixed offset, in decibels, that converts the adopted channel-gain normalization to the target path-loss convention for anchor $i$. It is constant with respect to the material parameters and may include known anchor-specific link-budget terms.

\subsection{Robust Objective and Physical Reparameterization}
\label{subsec:cal_loss}

Let $\mathcal M_{\mathrm{cal}}$ denote the set of calibrated material classes, let $M_{\mathrm{cal}}=|\mathcal M_{\mathrm{cal}}|$, and fix the material order $(m_1,\ldots,m_{M_{\mathrm{cal}}})$. For every $m\in\mathcal M_{\mathrm{cal}}$, let $u_m$ and $v_m$ be dimensionless unconstrained variables. We map them to admissible material domains and stack both the unconstrained vector $\mathbf z\in\mathbb R^{2M_{\mathrm{cal}}}$ and the normalized material vector $\boldsymbol\theta(\mathbf z)\in\mathbb R^{2M_{\mathrm{cal}}}$ in this order,
\begin{equation}
\begin{alignedat}{2}
&\mathbf z\;
&&=
[u_{m_1},v_{m_1},\ldots,u_{m_{M_{\mathrm{cal}}}},v_{m_{M_{\mathrm{cal}}}}]^{\mathsf T},\\
&\varepsilon_{r,m}(u_m)\;
&&=
\varepsilon_{\min}
+\!(\varepsilon_{\max}{-}\varepsilon_{\min})
\operatorname{sigmoid} (u_m), \\
&\sigma_m(v_m)\;&&=\sigma_{\mathrm{ref}}\exp(v_m),\\
&\boldsymbol\theta_m(u_m,v_m)\;
&&=\left[
\varepsilon_{r,m}(u_m),
\frac{\sigma_m(v_m)}{\sigma_{\mathrm{ref}}}
\right]^{\mathsf T},\\
&\boldsymbol\theta(\mathbf z)\;
&&=[\boldsymbol\theta_{m_1}^{\mathsf T},\ldots,
\boldsymbol\theta_{m_{M_{\mathrm{cal}}}}^{\mathsf T}]^{\mathsf T},\\
&\boldsymbol\theta(\mathbf z)\;
&&\in\mathbb R^{2M_{\mathrm{cal}}},
\end{alignedat}
\label{eq:reparam}
\end{equation}
In the stacking line, each block $\boldsymbol\theta_m$ abbreviates $\boldsymbol\theta_m(u_m,v_m)$ and the blocks follow the fixed material order. The quantities $\varepsilon_{\min}$ and $\varepsilon_{\max}$ are dimensionless lower and upper permittivity bounds, respectively, $\operatorname{sigmoid}(\cdot)$ is the logistic function, and $\sigma_{\mathrm{ref}}=1\,\mathrm{S/m}$. Thus, $\varepsilon_{r,m}$ is bounded, $\sigma_m$ is positive and expressed in S/m, and $\boldsymbol\theta$ uses the fixed dimensionless conductivity coordinate $\sigma_m/\sigma_{\mathrm{ref}}$. When conductivity is represented numerically in S/m, this choice of $\sigma_{\mathrm{ref}}$ makes the expression numerically identical to the shorthand $\sigma_m=\exp(v_m)$. The field replay recovers the physical conductivity as $\sigma_m=\sigma_{\mathrm{ref}}[\boldsymbol\theta_m(u_m,v_m)]_2$. In the reported experiment, $\mathcal M_{\mathrm{cal}}=\{\mathrm{marble}\}$ and $M_{\mathrm{cal}}=1$.

Let $\mathcal I_{\mathrm{tr}}\subseteq\mathcal I$ denote the training anchors, and let $\mathrm{PL}_i^{\mathrm{tar}}$ be the target path loss of anchor $i$, expressed in decibels. Define the residual $e_i(\mathbf z)=\widehat{\mathrm{PL}}_i(\boldsymbol\theta(\mathbf z))-\mathrm{PL}_i^{\mathrm{tar}}$. We minimize the Huber data objective~\cite{huber1964robust}
\begin{equation}
\mathcal{L}_{\mathrm{data}}(\mathbf{z})
=
\frac{1}{|\mathcal{I}_{\mathrm{tr}}|}
\sum_{i\in\mathcal{I}_{\mathrm{tr}}}
\rho_{\delta_{\mathrm H}}\!\left(e_i(\mathbf z)\right),
\label{eq:cal_huber}
\end{equation}
where the Huber penalty $\rho_{\delta_{\mathrm H}}$ is
\begin{equation}
\rho_{\delta_{\mathrm H}}(e)=
\begin{cases}
\frac{1}{2}e^2, & |e|\leq\delta_{\mathrm H},\\
\delta_{\mathrm H}\left(|e|-\frac{1}{2}\delta_{\mathrm H}\right),
& |e|>\delta_{\mathrm H}
\end{cases},
\label{eq:cal_huber_piecewise}
\end{equation}
where the generic residual $e$ and transition threshold $\delta_{\mathrm H}>0$ are expressed in decibels. Consequently, $\rho_{\delta_{\mathrm H}}$ and $\mathcal L_{\mathrm{data}}$ have a squared-decibel numerical scale. The quadratic region gives a smooth objective near the optimum, whereas the linear region limits the influence of large anchor errors. A weak prior regularizes directions that are poorly constrained by the anchors,
\begin{equation}
\mathcal{L}_{\mathrm{total}}(\mathbf z)
=
\mathcal{L}_{\mathrm{data}}(\mathbf z)
+\lambda_{\mathrm{prior}}
\left\|
\mathbf{z}-\mathbf{z}_0
\right\|_2^2,
\label{eq:cal_total}
\end{equation}
where $\mathbf z_0$ is the inverse image of the material-table initialization under~\eqref{eq:reparam}. This inverse is finite provided $\varepsilon_{\min}<\varepsilon_{r,m}^{(0)}<\varepsilon_{\max}$ and $\sigma_m^{(0)}>0$ for every $m\in\mathcal M_{\mathrm{cal}}$. We initialize $\mathbf z^{(0)}=\mathbf z_0$. The weight $\lambda_{\mathrm{prior}}\geq0$ has the same squared-decibel numerical scale as $\mathcal L_{\mathrm{data}}$. The optimizer therefore differentiates $\mathcal L_{\mathrm{total}}(\mathbf z)$ with respect to the unconstrained variables $\mathbf z$, propagating gradients through the mapping $\mathbf z\mapsto\boldsymbol\theta(\mathbf z)$ and the fixed-topology field replay. At iteration $q\in\{0,\ldots,Q-1\}$, where $Q$ is the number of optimization steps, Adam~\cite{kingma2015adam} performs the update
\begin{equation}
\mathbf{z}^{(q+1)}
=
\operatorname{Adam}
\left(
\mathbf{z}^{(q)},
\nabla_{\mathbf{z}}\mathcal{L}_{\mathrm{total}}\!\left(\mathbf z^{(q)}\right)
\right).
\label{eq:cal_update}
\end{equation}

\subsection{Gradient Verification and Identifiability}
\label{subsec:cal_loop}

We verify the reverse-mode gradient against a central finite difference at the same $\mathbf z$. For a coordinate index $a\in\{1,\ldots,2M_{\mathrm{cal}}\}$, let $\mathbf e_a$ denote the corresponding unit vector and let the dimensionless $\Delta_{\mathrm{FD}}>0$ denote the finite-difference step size. The two gradients are
\begin{equation}
\begin{aligned}
\mathbf{g}_{\mathrm{AD}}(\mathbf z)
&=\nabla_{\mathbf{z}}\mathcal{L}_{\mathrm{total}}(\mathbf z),\\
[\mathbf{g}_{\mathrm{FD}}\!(\mathbf z)]_a
&=\frac{\mathcal{L}_{\mathrm{total}}\!(\mathbf{z}\!+\!\Delta_{\mathrm{FD}}\mathbf{e}_a)
\!-\!\mathcal{L}_{\mathrm{total}}\!(\mathbf{z}\!-\!\Delta_{\mathrm{FD}}\mathbf{e}_a)}{2\Delta_{\mathrm{FD}}},
\end{aligned}
\label{eq:cal_grad_check}
\end{equation}
where $\mathbf{g}_{\mathrm{AD}}$ and $\mathbf{g}_{\mathrm{FD}}$ denote the automatic-differentiation and finite-difference gradients, respectively. Their relative discrepancy $e_{\mathrm{grad}}(\mathbf z)$ is
\begin{equation}
e_{\mathrm{grad}}(\mathbf z)
=\frac{\|\mathbf{g}_{\mathrm{AD}}(\mathbf z)-\mathbf{g}_{\mathrm{FD}}(\mathbf z)\|_2}
{\|\mathbf{g}_{\mathrm{FD}}(\mathbf z)\|_2+\epsilon_{\mathrm{grad}}},
\label{eq:cal_grad_error}
\end{equation}
where $\epsilon_{\mathrm{grad}}>0$ has the same squared-decibel numerical scale as the gradient norm and prevents division by zero. We further examine the training-anchor Jacobian $\mathbf J$ in the dimensionless material coordinates of~\eqref{eq:reparam},
\begin{equation}
\mathbf{J}
=
\frac{\partial
\bigl[\widehat{\mathrm{PL}}_i(\boldsymbol{\theta})\bigr]_{i\in\mathcal{I}_{\mathrm{tr}}}}
{\partial\boldsymbol{\theta}^{\mathsf T}},
\label{eq:cal_jacobian}
\end{equation}
where $\mathbf J\in\mathbb R^{|\mathcal I_{\mathrm{tr}}|\times2M_{\mathrm{cal}}}$ and $(\cdot)^{\mathsf T}$ denotes transpose. A material untouched by retained paths produces a zero column, whereas nearly collinear columns indicate that the corresponding parameters cannot be separated locally. Material coverage, Jacobian rank, and conditioning therefore determine whether the anchors constrain the selected variables. The fixed normalization in~\eqref{eq:reparam} makes the coordinate scale used for these diagnostics explicit; changing that scale would change the singular values and condition number, but not the rank. These diagnostics describe local sensitivity and do not establish global uniqueness.

\section{Dataset Format and Packaging}
\label{sec:packaging}
The dataset organizes all modalities under one hierarchical schema, so visual, geometric, and radio representations stay aligned in time and space. Each scenario is self-contained and uses the following directory structure.

{\footnotesize
\begin{verbatim}
scenario_name/
  |- model/      # 3D models and XML scene description
  |- config/     # Simulation configuration files
  |- snapshots/  # Rendered images and videos
  |- paths/      # Propagation path data (CSV/HDF5)
  |- heatmaps/   # Coverage maps (NPY/MAT)
  |- channel/    # CIR/CFR tensors (HDF5)
  `- metadata.json  # Scenario-level metadata
\end{verbatim}
}

This layout allows the user to load a single scenario and obtain its geometry, configuration, channels, and visualizations together. The dataset also remains extensible because new modalities map to new subdirectories under the same schema.

\section{Simulation and Analysis}
\label{sec:experiments}
DeepTelecom is evaluated through four complementary dimensions, including RIS-aware indoor\,/\,outdoor scene generation, UAV micro-Doppler modeling, material calibration on held-out synthetic anchors, and multimodal 3D localization.

\subsection{Scene Generation and RIS-Assisted Propagation}
\label{subsec:scenegen}
\begin{figure}[t]
    \centering
    \captionsetup[subfloat]{skip=3pt}
    \subfloat[Indoor RIS-assisted scene.]{%
        \includegraphics[width=0.85\columnwidth]{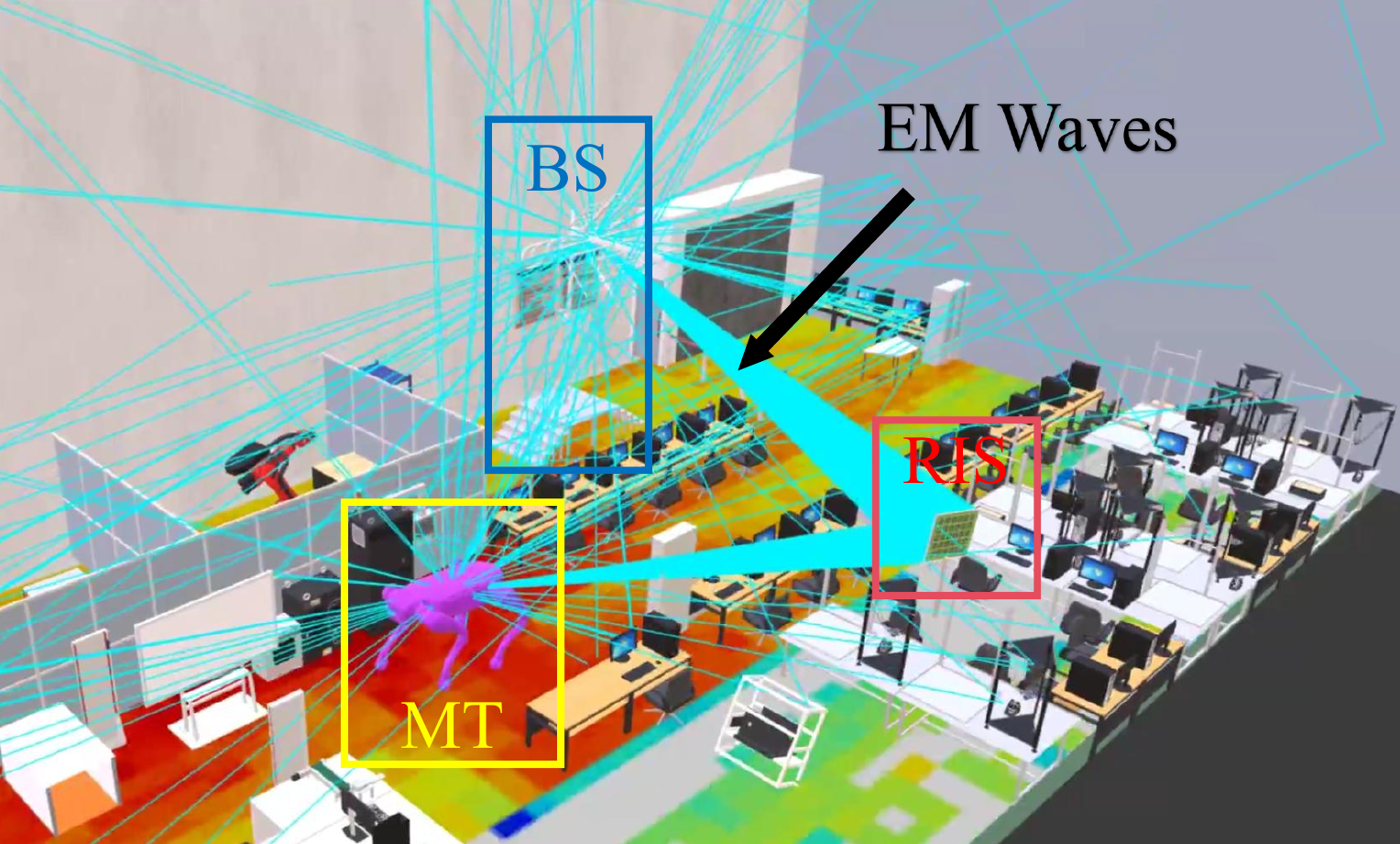}%
        \label{fig:indoor_vis}}
    \\[0.5ex]
    \subfloat[Outdoor RIS-assisted scene.]{%
        \includegraphics[width=0.85\columnwidth]{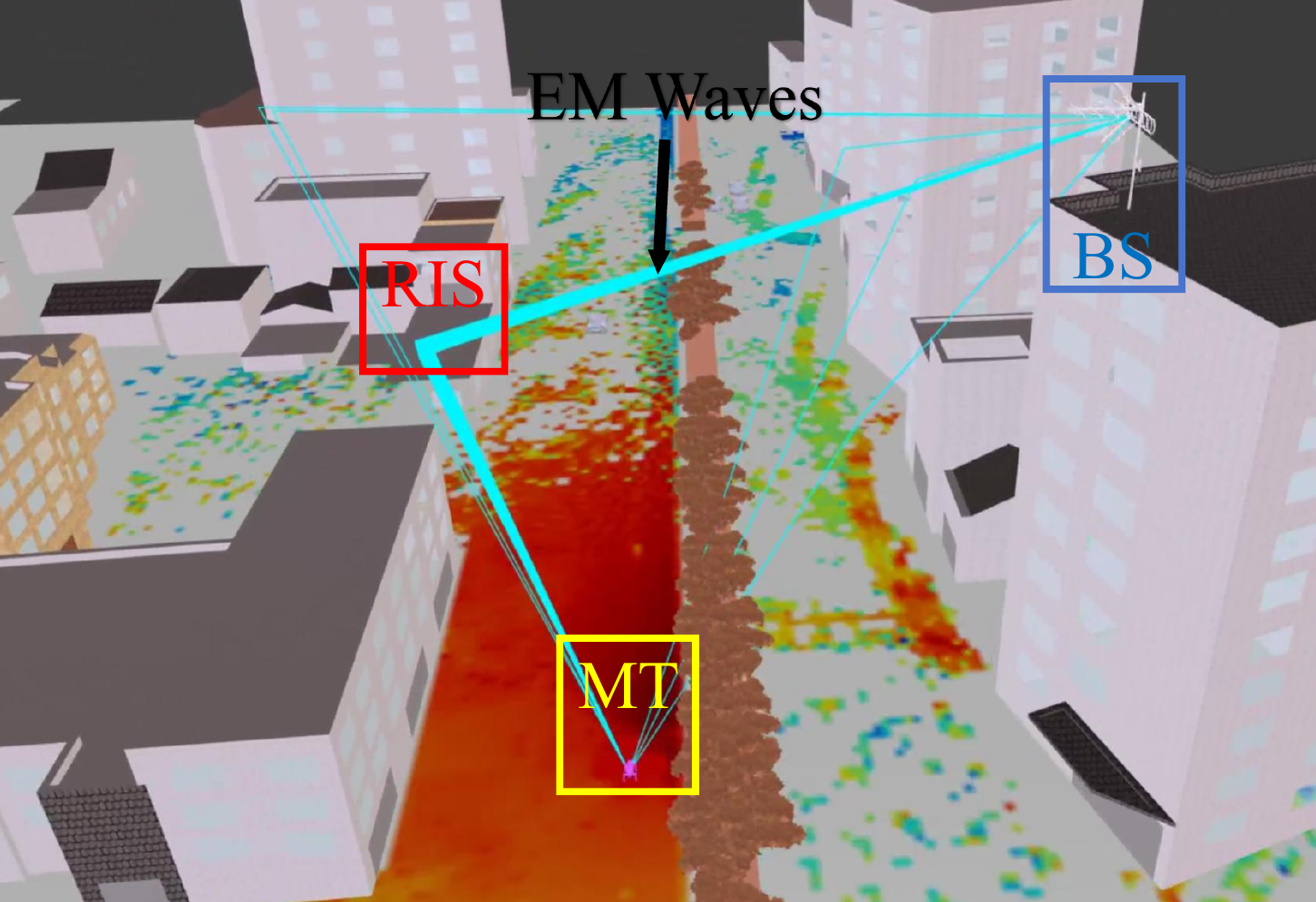}%
        \label{fig:outdoor_vis}}
    \caption{Material-aware LoD3 scenes with RIS-assisted propagation: (a) indoor and (b) outdoor.}
    \label{fig:scene_vis}
\end{figure}

Outdoor geospatial models and indoor point clouds are converted into a unified, material-aware LoD3 representation with consistent BS, MT, and RIS interfaces.
Fig.~\ref{fig:scene_vis}\subref{fig:indoor_vis} shows the indoor result. Walls, furniture, and partitions provide detailed interaction surfaces, and the traced paths connect the BS, RIS, and MT. The highlighted BS-RIS-MT path shows that RIS-assisted propagation is integrated into the indoor RT scene.
Fig.~\ref{fig:scene_vis}\subref{fig:outdoor_vis} shows the corresponding result in an outdoor street environment. Although the building geometry produces a different multipath structure, the same device and channel interfaces represent the RIS-assisted path. Together, the two cases demonstrate that heterogeneous scene sources converge to one simulation pipeline. 

\subsection{UAV Simulation}
\label{subsec:uav_sim}
\begin{table}[t]
\centering
\caption{User-configurable UAV and sampling parameters.}
\label{tab:uav_parameters}
\scriptsize
\setlength{\tabcolsep}{4pt}
\begin{tabular}{@{}>{\raggedright\arraybackslash}p{0.38\columnwidth}
                >{\raggedright\arraybackslash}p{0.56\columnwidth}@{}}
\hline\hline
Parameter & Setting used in this study \\
\hline
Carrier frequency $f_c$ & $28$\,GHz \\
Flight trajectory & B\'{e}zier curve \\
Translational speed $v$ & Constant $10$\,m/s \\
Pitch angle & $0^\circ$, $30^\circ$, and $45^\circ$ \\
Rotor frequency $f_{\mathrm{rot}}$ & $25$\,Hz, $\pm10\%$ per sample \\
Blade radius $R$ & $0.02$\,m, $\pm10\%$ per sample \\
Rotor and blade layout & Four rotors, two blades per rotor \\
Discretization of the full UAV & Three points per blade and one body point, $25$ in total \\
Single-blade reference & One blade-tip point \\
Camera/UAV sample rate $f_s$ & Shared and user-configurable \\
Snapshots per sample $N_{\mathrm{snap}}$ & User-configurable \\
Snapshot interval $\Delta t_{\mathrm{snap}}$ & User-configurable \\
Snapshot window $T_w$ & No longer than the local coherence time \\
\hline
\end{tabular}
\end{table}

\begin{figure}[t]
\centering
\includegraphics[width=0.98\columnwidth]{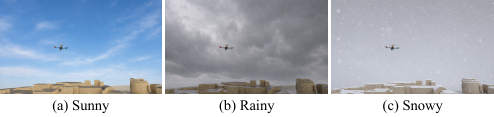}
\caption{Synchronized camera frames of the outdoor UAV scene under sunny, rainy, and snowy conditions.}
\label{fig:uav_scenario}
\end{figure}

We consider a quadrotor UAV in an outdoor ray-tracing scenario with a fixed rooftop BS operating at $28$\,GHz. The UAV follows a B\'{e}zier trajectory at a constant speed of $10$\,m/s. We evaluate pitch angles of $0^\circ$, $30^\circ$, and $45^\circ$. For each sample, we randomly vary the rotor frequency and blade radius within $\pm10\%$ of their nominal values. Table~\ref{tab:uav_parameters} summarizes the UAV and sampling variables provided to the user.

Let $b\in\{0,\ldots,N_{\mathrm{out}}-1\}$ index the $N_{\mathrm{out}}$ synchronized output samples. At time $t_b=b/f_s$, where $f_s$ is the shared output sampling rate, the renderer records one camera frame $\mathbf I_b$, and the simulator records the UAV state and radio data at the same timestamp. Thus, camera frames and UAV/radio samples are in one-to-one correspondence. Resolving the rotor motion within output sample $b$ requires the higher-rate complex snapshot vector $\mathbf s_b$. Let $n\in\{0,\ldots,N_{\mathrm{snap}}-1\}$ index the rotor snapshots, let $\Delta t_{\mathrm{snap}}$ denote their interval, and let $T_c(t_b)$ denote the local coherence time at $t_b$. Then,
\begin{equation}
\begin{aligned}
t_{b,n} &= t_b+n\Delta t_{\mathrm{snap}}, \\
\mathbf s_b
&=\left[s(t_{b,0}),\ldots,s(t_{b,N_{\mathrm{snap}}-1})\right]^{\mathsf T}
\in\mathbb C^{N_{\mathrm{snap}}}, \\
T_w &= (N_{\mathrm{snap}}-1)\Delta t_{\mathrm{snap}}\leq T_c(t_b).
\end{aligned}
\label{eq:uav_snapshot_array}
\end{equation}
The user can set both $N_{\mathrm{snap}}$ and $\Delta t_{\mathrm{snap}}$. Keeping all snapshots within one coherence interval makes the macroscopic propagation geometry locally quasi-static while resolving the rotor-induced phase variation used to generate the MDS.

Fig.~\ref{fig:uav_scenario} shows synchronized camera frames for the same UAV state under sunny, rainy, and snowy conditions. The sunny and snowy cases provide visual variation only. For rain, we additionally apply the ITU-R P.838-3 specific-attenuation model~\cite{iturp8383}, whose specific attenuation $\gamma_{\mathrm{rain}}$ is
\begin{equation}
\gamma_{\mathrm{rain}}=k_{\mathrm R}(f_c,p_{\mathrm{pol}})\,
r_{\mathrm{rain}}^{\alpha_{\mathrm R}(f_c,p_{\mathrm{pol}})}
\quad [\mathrm{dB/km}],
\label{eq:rain_specific_attenuation}
\end{equation}
where $f_c$ is expressed in GHz, $r_{\mathrm{rain}}$ is the rain rate in $\mathrm{mm/h}$, $p_{\mathrm{pol}}$ denotes the polarization and path-geometry configuration, and $k_{\mathrm R}(f_c,p_{\mathrm{pol}})$ and $\alpha_{\mathrm R}(f_c,p_{\mathrm{pol}})$ are the corresponding ITU-R coefficients. For scatterer $k\in\mathcal K$, let $d_k$ denote the one-way rain path length from the BS to that scatterer, expressed in kilometers. The rain-attenuated one-way channel is
\begin{equation}
h_k^{\mathrm{rain}}(t)=10^{-\gamma_{\mathrm{rain}}d_k/20}h_k(t).
\label{eq:rain_scaling}
\end{equation}
This positive real factor changes the echo amplitude but not its micro-Doppler frequency. Because the monostatic return in~\eqref{eq:md_echo} squares the one-way channel, the corresponding round-trip attenuation factor is $10^{-\gamma_{\mathrm{rain}}d_k/10}$. No RF snow-attenuation term is applied in the current experiment. Snow changes the rendered camera appearance only.

\begin{figure}[t]
\centering
\includegraphics[width=0.98\columnwidth]{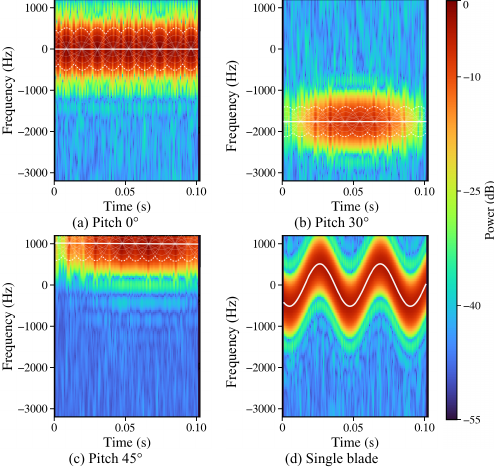}
\caption{Ray-traced micro-Doppler spectrograms at $28$\,GHz for (a) $0^\circ$, (b) $30^\circ$, and (c) $45^\circ$ pitch, and (d) the single-blade reference. Solid and dotted curves denote the predicted bulk and blade-tip Doppler shifts.}
\label{fig:uav_mds}
\end{figure}

\begin{figure}[t]
\centering
\includegraphics[width=0.98\columnwidth]{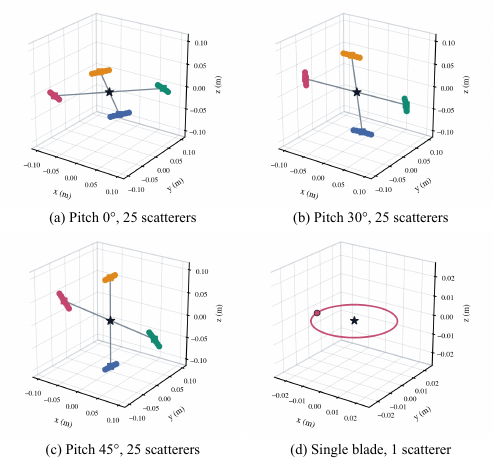}
\caption{UAV point-scatterer models for (a) $0^\circ$, (b) $30^\circ$, and (c) $45^\circ$ pitch, and (d) the single blade-tip reference.}
\label{fig:uav_scatterers}
\vspace{-0.9\baselineskip}
\end{figure}

The short-time Fourier transform of each snapshot vector in~\eqref{eq:uav_snapshot_array} produces the ray-traced spectrograms in Fig.~\ref{fig:uav_mds}. The solid and dotted curves are the independently computed bulk and blade-tip predictions, respectively. Although all full-airframe cases move at $10$\,m/s, their bulk shifts differ because the LoS projection of the velocity changes along the curved trajectory. The selected $0^\circ$ sample therefore has a near-zero radial velocity rather than zero flight speed. Pitch changes the LoS projection within the rotor plane, represented by $\cos\theta$, and hence the sideband envelope. At the nominal rotor frequency, two blades produce a spacing of $Bf_{\mathrm{rot}}\approx50$\,Hz. The single-blade case isolates the sinusoidal ridge predicted by~\eqref{eq:md_inst}.
Fig.~\ref{fig:uav_scatterers} shows the corresponding point-scatterer models. Each full UAV contains one body point and three radial points on each of the eight blades, giving $25$ scatterers. The single-blade reference retains one blade-tip point to isolate the elementary trajectory.

\subsection{Calibration Results}
\label{subsec:cal_results}

We evaluate calibration at $28$\,GHz in an outdoor scene using 48 low-altitude non-line-of-sight (NLoS) UAV anchors, divided into 24 training, 12 validation, and 12 held-out anchors. The path audit retains 97 propagation paths, with one to three paths per anchor, and records 142 interactions with the marble surface. All anchors exhibit nonzero sensitivity to the selected material parameters.

Target channels are generated using a designated marble setting that is hidden from the optimizer. We then reset the scene to its ITU-R initialization and optimize only the marble relative permittivity $\varepsilon_r$ and conductivity $\sigma$ from the 24 training path-loss targets using the objective in~\eqref{eq:cal_total}. Other materials, scattering, and link-budget terms remain fixed. The implementation uses Sionna RT 2.0.1, traces each anchor once, and replays the material-dependent fields for $Q=200$ updates of~\eqref{eq:cal_update}.

\begin{figure}[t]
\centering
\includegraphics[width=0.8\columnwidth]
{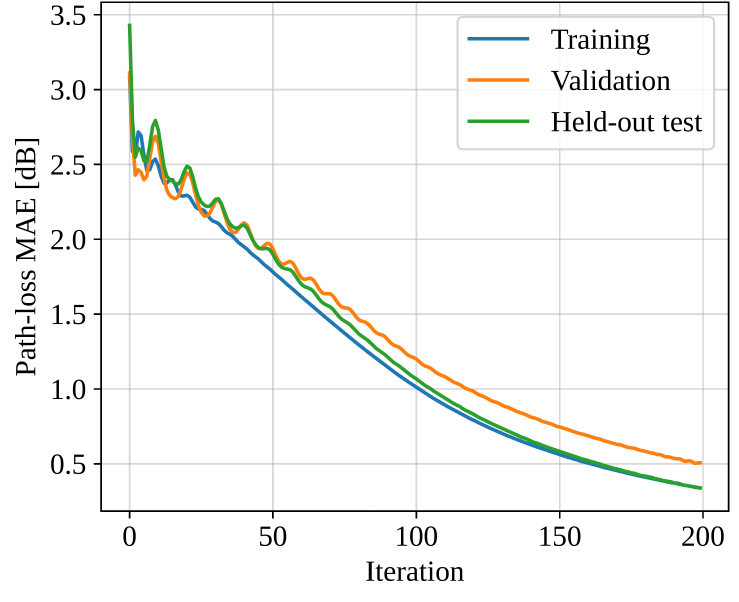}
\caption{Path-loss MAE over 200 calibration iterations at $28$\,GHz. Training anchors are used for optimization, whereas validation and held-out anchors are used to assess generalization.}
\label{fig:cal_conv}
\vspace{-0.9\baselineskip}
\end{figure}

Fig.~\ref{fig:cal_conv} shows that the path-loss mean absolute error (MAE) decreases steadily over 200 iterations and that the validation and held-out curves follow the training trend, indicating stable convergence and consistent generalization.
At the reported verification point, the relative discrepancy in~\eqref{eq:cal_grad_error} is $e_{\mathrm{grad}}=2.08\times10^{-3}$. The $24\times2$ training-anchor Jacobian in~\eqref{eq:cal_jacobian}, evaluated in the dimensionless coordinates of~\eqref{eq:reparam}, has rank two, numerical singular values $399.94$ and $6.36$, and a condition number of $62.9$. The two variables are therefore locally observable from these anchors, although their sensitivities are uneven and this local result does not imply a unique global solution.

On the 12 held-out anchors, calibration reduces the path-loss MAE from $3.429$\,dB to $0.343$\,dB and the root-mean-square error (RMSE) from $3.601$\,dB to $0.435$\,dB. For held-out anchor $i$, the relative absolute error (RAE) of the RMS delay spread in~\eqref{eq:delay_spread}, denoted by $\operatorname{RAE}_{\tau,i}$, is
\begin{equation}
\operatorname{RAE}_{\tau,i}
=
\frac{\left|\widehat{\tau}_{\mathrm{rms},i}
-\tau_{\mathrm{rms},i}^{\mathrm{tar}}\right|}
{\tau_{\mathrm{rms},i}^{\mathrm{tar}}+\epsilon_\tau},
\label{eq:cal_delay_rae}
\end{equation}
where $\widehat{\tau}_{\mathrm{rms},i}$ and $\tau_{\mathrm{rms},i}^{\mathrm{tar}}$ are the predicted and target RMS delay spreads for held-out anchor $i$, respectively, and $\epsilon_\tau>0$ is expressed in the same time unit as the delay spreads and stabilizes the ratio. Its mean decreases from $0.194$ to $0.066$, although delay spread is not included in the loss. This improvement arises because the learned material parameters reweight the retained paths and thereby change their relative contributions to channel dispersion.

\begin{figure}[t]
\centering
\includegraphics[width=0.9\columnwidth]
{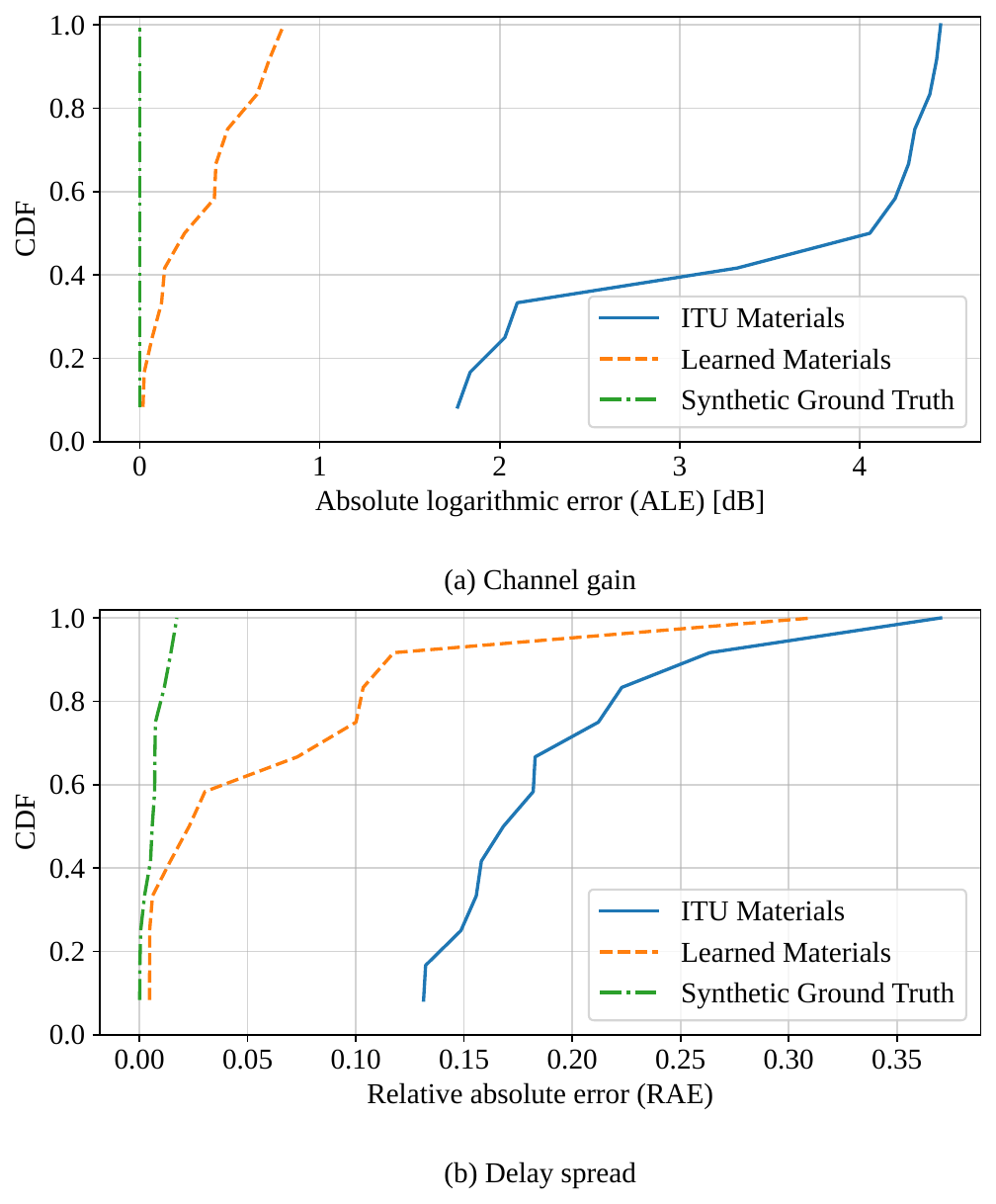}
\caption{Empirical CDFs of channel-gain ALE and RMS-delay-spread RAE over 12 held-out outdoor UAV anchors.}
\label{fig:cal_cdf}
\end{figure}

To characterize the full held-out error distributions, we use empirical cumulative distribution functions (CDFs) of the channel-gain absolute logarithmic error (ALE) and the RMS-delay-spread RAE.
As shown in Fig.~\ref{fig:cal_cdf} and Table~\ref{tab:calibration}, calibration shifts the held-out channel-gain ALE and RMS-delay-spread RAE distributions toward the synthetic ground truth, with a pronounced contraction of the channel-gain error tail. This distribution-wide improvement, together with the lower aggregate errors, demonstrates the effectiveness of the fixed-topology optimization.

\begin{table}[t]
\centering
\caption{Calibration performance on 12 held-out outdoor UAV anchors.}
\label{tab:calibration}
\scriptsize
\setlength{\tabcolsep}{4pt}
\begin{tabular}{@{}>{\raggedright\arraybackslash}p{0.31\columnwidth}
                >{\centering\arraybackslash}p{0.20\columnwidth}
                >{\centering\arraybackslash}p{0.20\columnwidth}
                >{\centering\arraybackslash}p{0.19\columnwidth}@{}}
\hline\hline
Material model & PL MAE [dB] & PL RMSE [dB] & RAE$_\tau$ \\
\hline
ITU-R initialization & 3.429 & 3.601 & 0.194 \\
Calibrated $\varepsilon_r$ and $\sigma$
& 0.343 & 0.435 & 0.066 \\
\hline
\end{tabular}
\end{table}

\subsection{Downstream Application}
\label{subsec:downstream}
To assess the downstream sensing utility of DeepTelecom, we consider device-free and noncooperative 3D localization of a passive UAV target via monostatic cellular integrated sensing and communication (ISAC). Here, passive describes the UAV target rather than the sensing infrastructure: a fixed rooftop BS transmits the sensing waveform and receives the monostatic echo, whereas the UAV neither communicates with the BS nor carries a localization device. We use $6{,}620$ samples from four UAV motion classes and divide them into training, validation, and test sets in an $8{:}1{:}1$ ratio, giving $5{,}296$, $662$, and $662$ samples, respectively. The real-valued network representation of the ordered complex echo vector $\mathbf s_b$ in~\eqref{eq:uav_snapshot_array} is
\begin{equation}
\mathbf X_b
=\begin{bmatrix}
\operatorname{Re}\{\mathbf s_b\}^{\mathsf T}\\
\operatorname{Im}\{\mathbf s_b\}^{\mathsf T}
\end{bmatrix}
\in\mathbb R^{2\times N_{\mathrm{snap}}}.
\label{eq:localization_echo_tensor}
\end{equation}
Thus, $\mathbf X_b$ is the ordered real--imaginary representation of the time-domain monostatic echo samples in $\mathbf s_b$, rather than either matrix-valued channel response $\mathbf H_{\mathrm{CIR}}$ or $\mathbf H_{\mathrm{CFR}}$. In this experiment, the task-level term CSI refers to $\mathbf X_b$ and $N_{\mathrm{snap}}=2048$. Each sample comprises $\mathbf X_b\in\mathbb R^{2\times2048}$, a camera image $\mathbf I_b\in\mathbb R^{3\times180\times320}$, and a path-loss image $\mathbf P_b\in\mathbb R^{3\times180\times320}$; the midpoint UAV coordinate $\boldsymbol p_b=(x_b,y_b,z_b)$ is the regression target. Both models use GroupNorm with eight groups and optimize the Smooth L1 loss on standardized coordinates. Table~\ref{tab:localization_parameters} lists the key training parameters.

The CSI-only model encodes the two real--imaginary rows of $\mathbf X_b$ in~\eqref{eq:localization_echo_tensor} with three one-dimensional convolutional neural network (CNN) blocks and passes the pooled feature directly to a multilayer perceptron (MLP) regression head. The multimodal model uses a one-dimensional channel encoder and independent two-dimensional encoders for the camera and path-loss images. Bidirectional feature-wise linear modulation (FiLM) exchanges channel and camera context, after which rank-16 bilinear fusion combines their features. The resulting embedding is concatenated with the path-loss feature and mapped to $\widehat{\boldsymbol p}_b\in\mathbb R^{3}$ by an MLP. The MDS spectrogram derived by the short-time Fourier transform is released with each sample but is not used by either model.

\begin{table}[t]
\centering
\caption{Key training parameters for UAV localization.}
\label{tab:localization_parameters}
\scriptsize
\setlength{\tabcolsep}{3pt}
\begin{tabular}{@{}>{\raggedright\arraybackslash}p{0.25\columnwidth}
                >{\raggedright\arraybackslash}p{0.25\columnwidth}
                >{\raggedright\arraybackslash}p{0.43\columnwidth}@{}}
\hline\hline
Variable & CSI-only & Multimodal \\
\hline
Input modalities & CSI & CSI, camera, path-loss \\
Channel tensor size & $2\times2048$ & $2\times2048$ \\
Image size & Not used & $3\times180\times320$ each \\
Optimizer & AdamW & AdamW \\
Learning rate & $3\times10^{-4}$ & $8\times10^{-4}$ \\
Weight decay & $10^{-4}$ & $10^{-3}$ \\
Batch size & $256$ & $64$ \\
\hline
\end{tabular}
\end{table}

\begin{figure}[t]
\centering
\captionsetup[subfloat]{skip=1.5pt}
\subfloat[Training and validation loss.]{%
    \includegraphics[width=0.9\columnwidth,trim=0 4pt 0 0,clip]{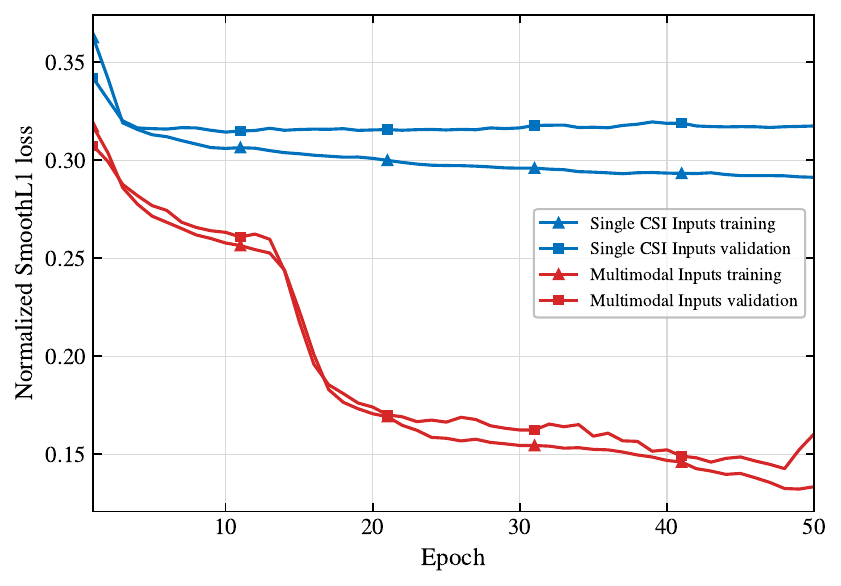}%
    \label{fig:localization_loss}}
\\[-0.5ex]
\subfloat[Localization-error CDF.]{%
    \includegraphics[width=0.9\columnwidth,trim=0 4pt 0 0,clip]{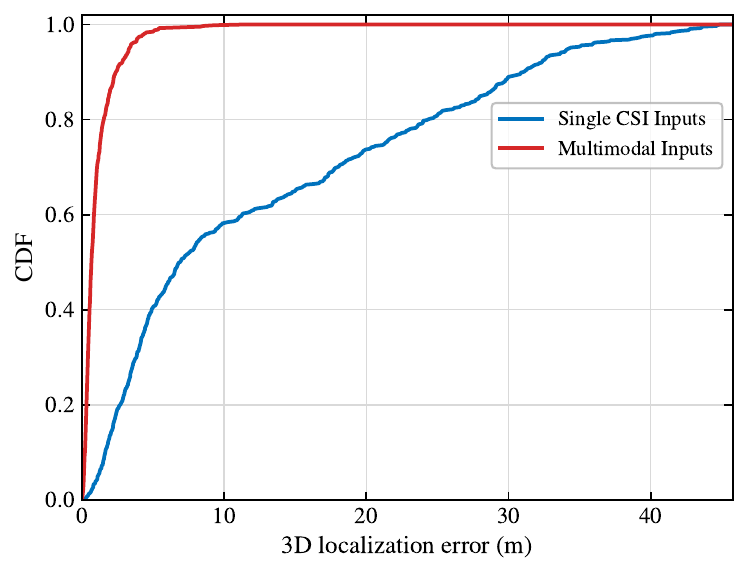}%
    \label{fig:localization_cdf}}
\caption{UAV localization performance for CSI-only and multimodal inputs: (a) normalized Smooth L1 training and validation losses and (b) CDFs of the 3D localization error. The multimodal input comprises CSI, camera images, and path-loss images.}
\label{fig:localization_results}
\vspace{-0.7\baselineskip}
\end{figure}

Fig.~\ref{fig:localization_results}\subref{fig:localization_loss} compares the normalized Smooth L1 losses over 50 epochs. At epoch 50, the multimodal validation loss is $0.160$, compared with $0.318$ for CSI alone, a $49.5\%$ reduction. The lower training and validation curves show that the added modalities provide a more informative representation for coordinate regression.
Fig.~\ref{fig:localization_results}\subref{fig:localization_cdf} shows the corresponding localization-error distributions. Adding camera and path-loss images reduces the median error from $6.96$ to $0.72$\,m and the 90th-percentile error from $31.05$ to $2.43$\,m. These visual features therefore reduce the spatial ambiguity of CSI-only localization.
This experiment establishes downstream utility within the synthetic UAV scene. Field validation and sim-to-real evaluation remain future work.
\section{Conclusion and Future Work}
\label{sec:conclusion}

DeepTelecom provides a unified material-aware LoD3 digital-twin framework and multimodal dataset for 6G wireless AI. It transforms coarse indoor\,/\,outdoor scenes into editable surface-level models, assigns frequency-dependent electromagnetic properties, and uses differentiable GPU RT to generate synchronized channel, visual, and weather-conditioned products for scenes containing RISs, MTs, and dynamic UAVs. The UAV model further captures translational velocity, pitch angle, and rotor-induced micro-Doppler. In a $28$\,GHz outdoor study, fixed-topology calibration reduces held-out path-loss MAE from $3.429$\,dB to $0.343$\,dB and RMS-delay-spread RAE from $0.194$ to $0.066$. For downstream 3D localization, combining CSI with camera and path-loss images reduces the 90th-percentile error from $31.05$\,m to $2.43$\,m. Together, these results demonstrate channel-level calibration and multimodal learning within the evaluated synthetic scenes. Users can generate customized data for their own scenes and system configurations to support a broad range of applications, including UAV identification, localization, tracking, beam prediction, and channel estimation.

Future work will establish a bidirectional loop between DeepTelecom and its physical counterpart. Live wireless and sensing measurements will update scene, material, channel, and UAV states online, while DeepTelecom will return channel predictions and guidance for beam selection, RIS configuration, and UAV trajectory control. This loop will support continual calibration, field validation, and sim-to-real transfer across diverse materials and trajectories. We will also enrich signal-level representations and sensing modalities and refine dataset generation through improved scene geometry, environmental rendering, and models of object dynamics, moving scatterers, and rotating platforms.

\small
\bibliographystyle{IEEEtran}
\bibliography{bib}

\end{document}